\documentclass[aps,prl,twocolumn,amsfonts,amssymb,amsmath,10pt,showpacs,showkeys,superscriptaddress]{revtex4-1}
\usepackage{graphicx}
\usepackage{natbib}
\usepackage{dcolumn}%
\usepackage{bm}%
\usepackage[usenames]{color}
\begin{document}
\title{Einstein-de Haas Effect in a Dipolar Fermi Gas}
\date{\today}

 \author{Ulrich~Ebling}
  \email[Electronic address:]{ebling@cat.phys.s.u-tokyo.ac.jp} 
 \affiliation{Department of Physics, University of Tokyo, 7-3-1 Hongo, Bunkyo-ku, Tokyo 113-0033, Japan}

 \author{Masahito Ueda}
 \affiliation{Department of Physics, University of Tokyo, 7-3-1 Hongo, Bunkyo-ku, Tokyo 113-0033, Japan} 
 \affiliation{RIKEN Center for Emergent Matter Science (CEMS), Wako, Saitama 351-0198, Japan} 

\begin{abstract} 
We show that an analogue to the classical Einstein-de Haas effect can appear in ultracold dipolar Fermi gases. The anisotropic nature of dipole-dipole interactions can lead to a transfer of magnetization into orbital angular momentum. Remarkably, distinct from a Bose-Einstein condensate, this transfer is accompanied by twisting motion, where individual spin components rotate in opposite directions with larger orbital angular momenta than the full system, possibly leading to easier experimental observation of the effect. This feature is induced by the deformation of the Fermi surface and the direction of the twisting motion can be controlled by an s-wave scattering length or external magnetic field, possibly providing a method of measuring scattering lengths of strongly dipolar atomic species.  
\end{abstract}

\maketitle
Over the last decade, experimental progress in optical cooling and trapping techniques has led to the achievement of quantum degeneracy of  strongly dipolar atoms, from the pioneering work on Chromium \cite{Griesmaier2005,Beaufils2008} to more recent experiments with Dysprosium and Erbium \cite{Lu2011,Aikawa2012,Lu2012,Aikawa2013}. Interactions in these systems are not only long-ranged, but also anisotropic; they depend on the relative position and orientation of atoms \cite{Lahaye2009,Baranov2012}. Therefore, they intrinsically combine spin and orbital degrees of freedom. For instance a dipolar Fermi gas deforms its Fermi surface along the spin polarization axis \cite{Miyakawa2008,Aikawa2014}. On the orbital side, due to being both attractive and repulsive, dipolar Bose-Einstein condensates (BEC) are known to have complex stability properties \cite{Koch2008,Lahaye2008}, observable recently in the form of droplets stabilized by quantum fluctuations \cite{FerrierBarbut2016,Kadau2016,Schmitt2016,Waechtler2016,Baillie2016}. On the spinor side, dipole-dipole interactions, unlike s-wave scattering, do not conserve total magnetization \cite{Kawaguchi2012} and are expected to lead to intricate ground state phase diagrams \cite{Ueda2002,Barnett2006} as well as the ability to transfer magnetization into orbital angular momentum, analogous to the Einstein-de Haas (EdH) effect \cite{Kawaguchi2006,Santos2006,Gawryluk2007}.

The experimental vindication of the genuine EdH effect in dipolar BECs has been elusive, despite substantial experimental progresses. In this Letter, we propose a dipolar Fermi gas as a better candidate, because of a special feature of the Fermi system: the deformation of the Fermi surface. We demonstrate that this effect leads to an additional twisting motion of the Fermi system on top of the center-of-mass orbital angular momentum that is converted from initial magnetization by the EdH effect. Using numerical simulations of the dynamics of a simple two-component, two-dimensional (2D) Fermi gas, we show that both spin components $\lbrace\uparrow ,\downarrow\rbrace$ acquire angular momenta in opposite directions, with individual angular momenta considerably larger than that induced by the EdH effect $|L_\uparrow-L_\downarrow|\gg|L_\uparrow+L_\downarrow|$. Our results show that this twisting motion arises from Fermi surface deformation during the time evolution, which explains the absence of this effect in dipolar BEC \cite{Kawaguchi2006}. Therefore, this excess of the angular momentum of individual spin components can make the EdH effect easier to observe in a Fermi gas. 
In addition, manipulation of either an s-wave scattering length or an external magnetic field can reverse this twisting motion and change the sign of $L_\uparrow-L_\downarrow$, while leaving the orbital angular momentum of the full system $|L_\uparrow+L_\downarrow|$ unchanged. The sensitivity to the scattering length could help resolve the outstanding issue of measuring scattering lengths in Dy \cite{Tang2015} and Er.

We consider a dipolar Fermi gas of mass $m$ with magnetic dipole moment $\mu$, confined in a 2D geometry. The system is confined in the $z$-direction by a strong harmonic confinement $\omega_z$ such that all atoms occupy the harmonic oscillator ground state with respect to the $z$-axis. In the $x-y$ plane, we assume an isotropic harmonic trap $\omega=\omega_x=\omega_y\ll\omega_z$ so that the gas will form a radially symmetric disk. We study the case of zero or weak magnetic field $B$ perpendicular to the 2D disk. Therefore, the dipoles are not necessarily fully polarized and can vary in space and time. In this Letter, for the sake of simplicity of our theory and numerical simulations, we assume a 2-component Fermi gas with spin states $\lbrace\uparrow ,\downarrow\rbrace$, which in addition to dipole forces may also interact via the s-wave scattering length $a$, where the system is not fully polarized. 

The Hamiltonian of our system is
\begin{align}
\label{eq:1}
\hat H=&\int d^2r\sum_{k=\uparrow,\downarrow}\hat\psi_k^\dagger(\vec r)\left[\frac{-\hbar^2\nabla^2}{2m}+\frac{m\omega^2\vec r^2}{2}+Pk\right]\hat\psi_k(\vec r)\nonumber\\
&+g_\text{2D}\int d^2r\hat\psi_\uparrow^\dagger(\vec r)\hat\psi_\downarrow^\dagger(\vec r)\hat\psi_\downarrow(\vec r)\hat\psi_\uparrow(\vec r)\nonumber\\
+\frac12\int &d^2r d^2r'\sum_{klmn}\hat\psi_k^\dagger(\vec r)\hat\psi_l^\dagger(\vec r')V_{klmn}(\vec r-\vec r')\hat\psi_m(\vec r')\hat\psi_n(\vec r),
\end{align}
where $P=g\mu_B B$ is the Zeeman splitting, $g_\text{2D}=2\sqrt{2\pi}\hbar^2a/l_z m$ describes s-wave scattering in the 2D case with $l_z=\sqrt{\hbar/m\omega_z}$ characterizing the width of the system in the $z$-direction. The DDI potential is given by
\begin{align}
\label{eq:2}
V_{klmn}(\vec r)=&\frac{c_d\sqrt{2}}{\sqrt{\pi}l_z^5}\left(V_1(r)\left[\sigma_{kl}^z\sigma_{mn}^z-\tfrac14\sigma_{kl}^+\sigma_{mn}^-\right.\right.\nonumber\\
&\left.\left. -\tfrac14\sigma_{kl}^-\sigma_{mn}^+\right]-\tfrac14(x-iy)^2V_2(r)\sigma_{kl}^+\sigma_{mn}^+\right.\nonumber\\
&\left. -\tfrac14(x+iy)^2V_2(r)\sigma_{kl}^-\sigma_{mn}^- \right).
\end{align}
Its strength is determined by $c_d=\mu_0\mu^2/4\pi$ ($\mu_0$ is the permeability of vacuum), while its long-range behavior in 2D is given by
\begin{align}
\label{eq:3}
V_1(r)&=e^\frac{r^2}{4l_z^2}\left[(2l_z^2+r^2)K_0(\tfrac{r^2}{4l_z^2})-r^2K_1(\tfrac{r^2}{4l_z^2})\right],\\
\label{eq:4}
V_2(r)&=e^\frac{r^2}{4l_z^2}\left[K_0(\tfrac{r^2}{4l_z^2})+\left(\tfrac{2l_z^2}{r^2}-1\right)K_1(\tfrac{r^2}{4l_z^2})\right],
\end{align}
where $K_0$ and $K_1$ denote modified Bessel functions \footnote{See Supplemental Material for the derivation of the 2D interaction terms.}. 

\begin{figure}[t]
\centering
\includegraphics[width=8.6cm]{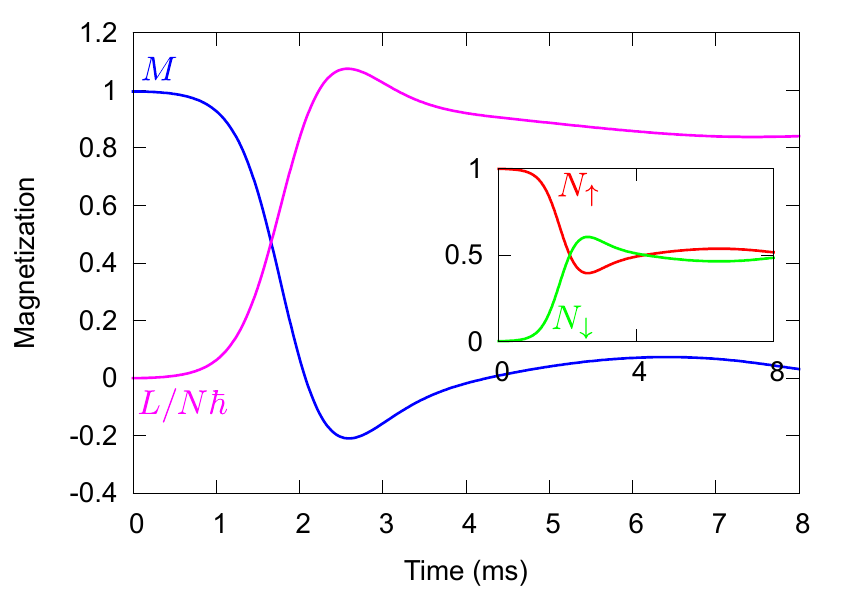}
\caption{Einstein-de Haas effect: Conversion of magnetization $N_\uparrow-N_\downarrow$ (blue) into orbital the angular momentum (magenta) due to the DDI for $B=0$ and $a=0$. The inset shows the time evolution of the spin populations of $N_\uparrow$ (red) and $N_\downarrow$ (green). Note that the system not only fully demagnetizes, but temporarily reaches negative magnetization around $2.6\,\text{ms}$.}
\label{fig1}
\end{figure}

We treat the dynamics of the system in the Hartree-Fock approximation, by following the time evolution of the one-body Wigner function
\begin{equation}
\label{eq:5}
 W_{mn}(\vec r,\vec p)=\int d^2r' e^{i\vec r'\vec p/\hbar}\left\langle\hat\psi_m^\dagger(\vec r-\vec r'/2)\hat\psi_n(\vec r+\vec r'/2)\right\rangle,
\end{equation}
which is the phase-space representation of the single-particle density matrix \cite{Goral2001,Miyakawa2008,Endo2010,Ebling2011,Babadi2012}. In the semi-classical approximation, in the collisionless regime, the dynamics of the system can be described by a collisionless Boltzmann-Vlasov equation (BVE)
\begin{align}
\label{eq:6}
\frac{d}{dt}&W(\vec r,\vec p)=\left(-\frac{\vec p}{m}\cdot\nabla_x+m\omega^2\vec r\cdot\nabla_p\right)W(\vec r,\vec p)\nonumber\\
&+\frac{1}{i\hbar}\left(\left[W(\vec r,\vec p),i\hbar p\sigma_z+U^a(\vec r)+U(\vec r)-\tilde U(\vec r,\vec p)\right]\right)\nonumber\\
&-\frac12\left(\left\lbrace\nabla_p W(\vec r,\vec p),\cdot\nabla_r \left(U^a(\vec r)+U(\vec r)-\tilde U(\vec r,\vec p)\right)\right\rbrace\right.\nonumber\\
&\left.+\left\lbrace\nabla_r W(\vec r,\vec p),\cdot\nabla_p\tilde U(\vec r,\vec p)\right\rbrace\right).
\end{align}
Equation~(\ref{eq:6}) involves single-particle contributions from the kinetic, trap and Zeeman energy, as well as contact interactions and two contributions from dipole-dipole interactions, the direct term that operates in real space and the exchange term in momentum space \footnote{For short-range interactions, the direct and exchange contributions are identical and result in a factor of 2.}. The elements of the mean-field potentials in Eq.~(\ref{eq:6}) are given by
\begin{align}
\label{eq:7}
U_{ij}^a(\vec r)&=g_\text{2D}\int\frac{d^2q}{(2\pi\hbar)^2}\left(\delta_{ij}\text{Tr}W(\vec r,\vec q)-W_{ij}(\vec r,\vec q)\right),\\
\label{eq:8}
U_{ij}(\vec r)&=\int\frac{d^2qd^2r'}{(2\pi\hbar)^2}\sum_{kl}V_{jikl}(\vec r-\vec r')W_{kl}(\vec r',\vec q),\\
\label{eq:9}
\tilde U_{ij}(\vec r,\vec p)&=\int\frac{d^2q}{(2\pi\hbar)^2}\sum_{kl}\tilde V_{jlki}(\vec q-\vec p)W_{kl}(\vec r,\vec q),
\end{align}
where $\tilde V$ denotes the momentum representation of the dipole-dipole interaction potential. We numerically integrate Eq.~(\ref{eq:6}) using the MacCormack method, where we work in Fourier space to calculate the mean-field potentials in Eqs.~(\ref{eq:8}) and (\ref{eq:9}) and cope with difficulties arising from the singularity of $V_1(r)$ at $r=0$ by using techniques applied to a dipolar BEC in Ref.~\cite{Ronen2006}.

We assume our system to be prepared initially in a strong magnetic field perpendicular to the $x-y$ plane, such that it is fully polarized along the $z$-axis, with $N_\downarrow=\int d^2rd^2pW_{\downarrow\downarrow}(\vec r,\vec p,t=0)=0$. In a 2D geometry, this means that the DDI is isotropic and repulsive, and we use the corresponding equilibrium distribution derived in Ref.~\cite{Babadi2012}. 
Next, the magnetic field is ramped down to very low values, such that the dipoles can evolve in time. In our numerical simulations, we assume the trap parameters to be $\omega=2\pi\times 84\,\text{Hz}$ and $\omega_z=2\pi\times 47\,\text{kHz}$ with $N=500$ and $T=0.2T_F$. We choose {}\textsuperscript{161}Dy with $m=160.93\,\text{u}$, $\mu=9.93\,\mu_\text{B}$ and $g=1.243$ as atomic species, which for simplicity we treat as a two-component Fermi gas.

For the simplest case of $B=0$ and $a=0$, the time-evolution of the two spin components is depicted in the inset of Fig.~\ref{fig1}. As shown there, from an initially fully polarized spin state with $N_\uparrow=N$, the Fermi gas evolves into an unpolarized state with $N_\downarrow=N_\uparrow$. This magnetization $M=N_\uparrow-N_\downarrow$ is fully converted into the orbital angular momentum $L=L_\uparrow+L_\downarrow$, where
\begin{equation}
\label{eq:11}
L_m=\int\frac{d^2rd^2p}{(2\pi\hbar)^2}(xp_y-yp_x)W_{mm}(\vec x,\vec p).
\end{equation}
 We note that, compared to the case of a dipolar BEC \cite{Kawaguchi2006}, demagnetization is more complete in our Fermi system, which quickly reaches zero magnetization \footnote{We verified that the complete demagnetization we observe is not just due to the stronger dipole moment of Dysprosium compared to Chromium. Using the parameters of Chromium in our simulations leads to slower dynamics but with the same degree of demagnetization.}.

\begin{figure}[t]
\centering
\includegraphics[width=8.6cm]{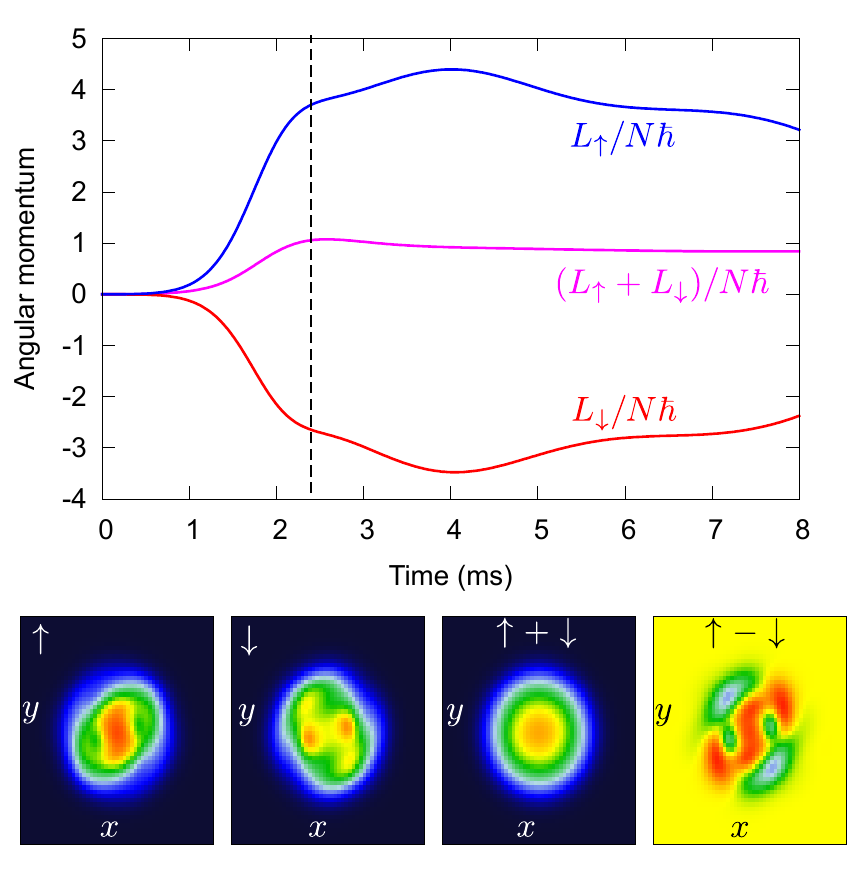}
\caption{Top: Twisting motion of a dipolar Fermi gas. Both spin components (red, blue) acquire large angular momenta in opposite directions. The angular momentum of the entire system (magenta, same as in Fig.~\ref{fig1}) is much smaller than the individual angular momenta. Bottom: Density plots at $t=2.6\,\text{ms}$ (dashed line) of $N_{mn}(x,y)=\int d^2p W_{mn}(\vec r,\vec p)$ for the spin components $\uparrow$, $\downarrow$, the sum and their difference. Spatial patterns associated with the generated angular momentum are expected to be more visible in an experiment for individual spin components or their differences.}
\label{fig2}
\end{figure}

\begin{figure}[t]
\centering
\includegraphics[width=8.6cm]{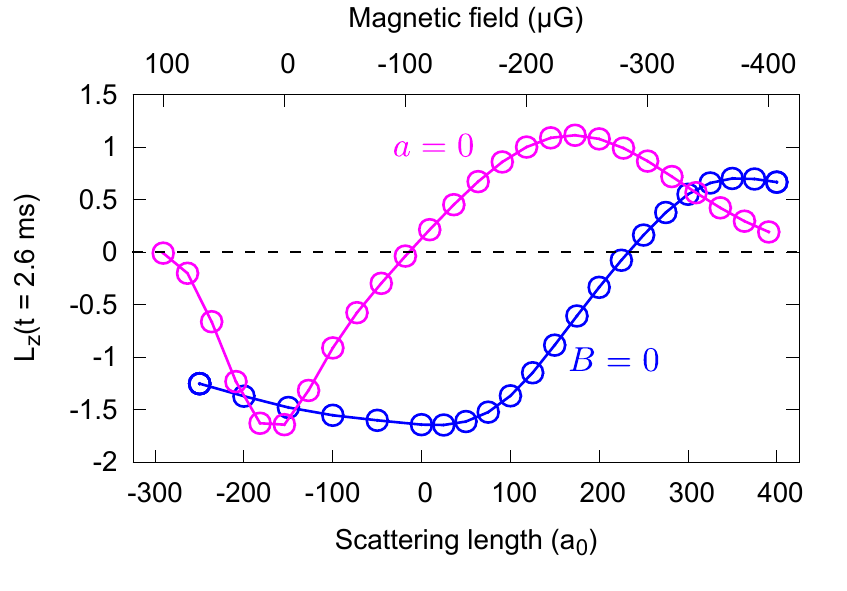}
\caption{Reversal of the twisting motion: Beyond certain values for $a$ (blue) and $B$ (magenta), the spin components reverse their sense of rotation. Shown is the relative angular momentum $L_z$ at $t=2.6\,\text{ms}$. The turning points $L_z=0$ (dashed line) correspond to the values at which $E_\text{kin}+E_\text{trap}$ remains approximately constant.}
\label{fig3}
\end{figure}

A surprising effect reveals itself when we decompose the total orbital angular momentum induced by the EdH effect into the the two spin components. We show the time evolution of $L_\uparrow$, $L_\downarrow$ and $L_\uparrow+L_\downarrow$ in Fig.~\ref{fig2}. We observe that the angular momentum is not only created in the $\downarrow$-component as expected, but in both components such that the $\downarrow$-component exhibits an excess of angular momentum, while the $\uparrow$-component rotates in the opposite direction. Even more remarkable is the fact that the magnitudes of both $L_\uparrow$ and $L_\downarrow$ are considerably larger than that of the total angular momentum as shown in the upper panel of Fig.~\ref{fig2}. In the lower panels of Fig.~\ref{fig2}, we plot density profiles of both spin components, their sum and difference. It is evident from these plots that in an experiment, finding the signature spatial patterns of the EdH effect is more likely to be successful by in-situ imaging individual spin components, rather than the full system, thanks to the additional twisting motion of the spin components we find in Fermi systems.

To understand the physical origin of this peculiar twisting motion, we take a look at the time evolution of the expectation value of orbital angular momentum $L$. We split the Wigner function and other single-particle quantities into a scalar and a vector part $W_{mn}(\vec r,\vec p)=W_0(\vec r,\vec p)\delta_{mn}+\vec W(\vec r,\vec p)\cdot\vec\sigma$. The scalar part of Eq.~(\ref{eq:11}), $L_0=L_\uparrow+L_\downarrow$, corresponds to the total angular momentum induced by the EdH effect. We find, that its time evolution is described by the subleading term (i.e. the anti-commutator) in the Boltzmann-Vlasov equation (\ref{eq:6}). However, the twisting motion, captured by the time evolution of $L_z=L_\uparrow-L_\downarrow$, is dominated by the commutator term in Eq.~(\ref{eq:6}), which explains its greater magnitude. We find that for short times, $L_z$ evolves according to
\begin{align}
\label{eq:12}
\frac{d}{dt}L_z\approx & \frac{\sqrt2c_dM_xM_y}{\sqrt{\pi}4l_z^5}\nonumber\\
&\times\int\frac{d^2rd^2r'd^2p}{(2\pi\hbar)^2}\left(xp_y-yp_x\right)V_2(\vec r-\vec r')\nonumber\\
&\times\left[(x-x')^2-(y-y')^2\right]N_0(\vec r')W_0(\vec r,\vec p),
\end{align} 
where $M_{x,y}$ denotes the transverse magnetization. So, in order for a twisting motion with $|L_z|>L_\downarrow$ to occur, a non-zero transverse magnetization must be present, and the total phase-space distribution of the system must be anisotropic. Both of these effects are related, since the transverse magnetization in turn is the source of the Fermi surface deformation \cite{Miyakawa2008}. Thus, small amounts present at $t=0$ or later during the demagnetization can lead to a twisting motion of the dipolar Fermi gas. The dependence of Eq.~(\ref{eq:12}) on an anisotropic real and momentum space distribution explains why such an effect has not been predicted for a dipolar BEC.
\begin{figure}[t]
\centering
\includegraphics[width=8.6cm]{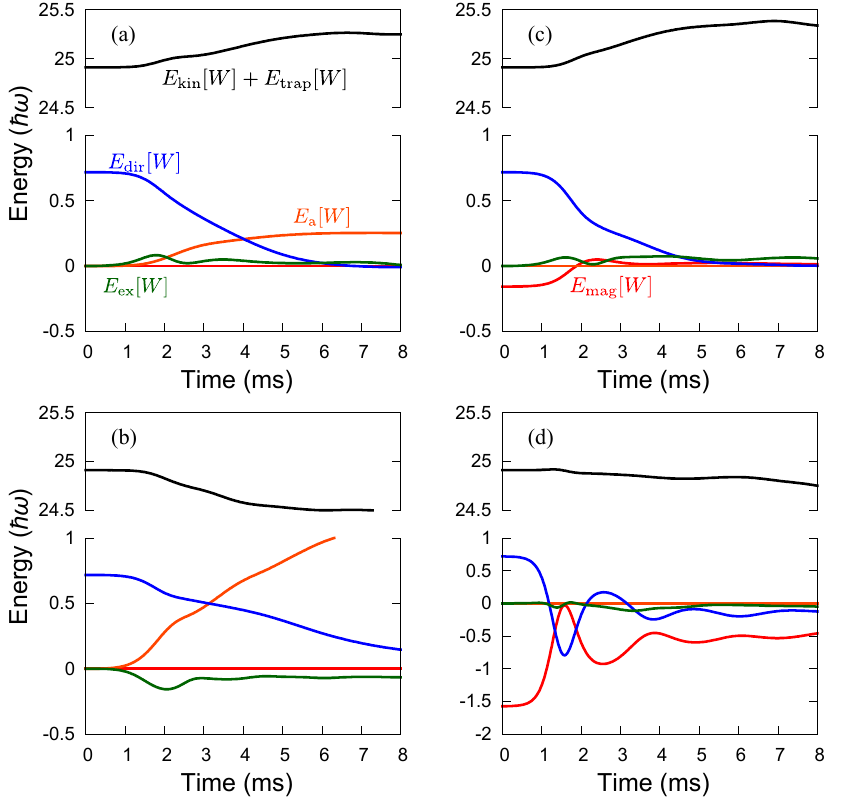}
\caption{Time evolution of different energy contributions. In the left column, $B=0$ and $a=50\,a_0$ (a) and $a=300\,a_0$ (b), on both sides of the sign change in Fig.~\ref{fig3}. Likewise, in the right column, $a=0$ and $B=-20\,\mathrm{\mu G}$ (c) and $B=-200\,\mathrm{\mu G}$ (d). The reversal of the twisting motion that accompanies the EdH effect coincides with a decrease/increase of the sum of kinetic and trap energies (black line).}
\label{fig4}
\end{figure}

Let us now turn our attention to another intriguing feature of the twisting motion that accompanies the EdH effect in dipolar Fermi gases. So far, we have considered a system without s-wave scattering and magnetic field $B=0$. In Fig.~\ref{fig3}, we show the relative orbital angular momentum of both spin components at $t=2.6\,\text{ms}$, approximately when the magnetization has reaches its minimum, for various values of $B$ and $a$. While the total angular momentum remains largely unchanged, apart from magnetic fields where the EdH is fully suppressed, the counter-rotating spin components actually change their relative motion at non-zero values of both $B$ and $a$. In other words, the $\downarrow$ spin component starts rotating in opposite direction as would be expected from the change of magnetization accompanying an interaction process $\lbrace\uparrow,\uparrow\rbrace\rightarrow\lbrace\downarrow,\downarrow\rbrace$ that generates orbital angular momentum. 
To investigate this feature, we take a look at the time evolution of the different energy contributions. We calculate the expectation value of the individual parts of the Hamiltonian Eq.~(\ref{eq:1}) in the Hartree-Fock approximation \cite{Miyakawa2008,Sogo2009} and obtain the normalized energy functionals
\begin{align}
\label{eq:13}
E_\text{kin}[W]&=\int \frac{d^2rd^2p}{(2\pi\hbar)^2N}\frac{p^2}{2m}W_0(\vec r,\vec p),\\
\label{eq:14}
E_\text{trap}[W]&=\int \frac{d^2rd^2p}{(2\pi\hbar)^2N}\frac{m\omega^2\vec r^2}{2}W_0(\vec r,\vec p),\\
\label{eq:15}
E_\text{mag}[W]&=\int \frac{d^2rd^2p}{(2\pi\hbar)^2N}PW_z(\vec r,\vec p),\\
\label{eq:16}
E_\text{a}[W]&=\int \frac{d^2rd^2p}{(2\pi\hbar)^2 2N}U^a(\vec r)W(\vec r,\vec p),\\
\label{eq:17}
E_\text{dir}[W]&=\frac12 \int \frac{d^2rd^2p}{2N(2\pi\hbar)^2}U(\vec r)W(\vec r,\vec p),\\
\label{eq:18}
E_\text{ex}[W]&=-\frac12\int \frac{d^2rd^2p}{2N(2\pi\hbar)^2}\tilde U(\vec r,\vec p)W(\vec r,\vec p).
\end{align}
These denote respectively the kinetic (\ref{eq:13}), trap (\ref{eq:14}) and Zeeman (\ref{eq:15}) energies as well as the short-range interaction energy (\ref{eq:16}) and the direct (\ref{eq:17}) and exchange (\ref{eq:18}) parts of the DDI. In Fig.~\ref{fig4}, we show the time evolution of these energy contributions for values of $a$ and $B$ on both sides of the sign change of the twisting motion (see Fig.~\ref{fig3}). The DDI is initially repulsive, hence $E_\text{dir}[W(t=0)]>0$. For $a=B=0$, total energy conservation means that $E_\text{kin}[W]+E_\text{trap}[W]$ must increase. However, $a$ and $B$ can be tuned such that $E_\text{kin}[W]+E_\text{trap}[W]$ must decrease, either if $E_\text{dir}[W(t=0)]<E_\text{a}[W(t\rightarrow\infty)]$ or $ E_\text{dir}[W(t=0)]+E_\text{mag}[W(t=0)]<0$, where we denote by $W(t\rightarrow\infty)$ the demagnetized state after the short-term EdH dynamics. This coincides with the sign change of $L_z$ depicted in Fig.~\ref{fig3}.

The reason for this is the Fermi surface deformation effect \cite{Miyakawa2008,Aikawa2014}. Sogo et al. showed, that an increase/decrease of the kinetic energy determines whether a Fermi gas expands or contracts along the transverse spin direction when undergoing the Fermi surface deformation \cite{Sogo2009}. A decrease in kinetic energy leads to a contraction along the spin polarization axis, while an increase leads to an elongation. In turn, this behavior determines the sign of the relative rotation of both spin components in Eq.~(\ref{eq:12}). This explains why an increase of an applied magnetic field or an s-wave scattering length beyond a critical value reverses the twisting motion of the Fermi gas due to an overall reduction of kinetic and trap energies.

In conclusion, we have shown that dipolar Fermi gases can exhibit transfer of spin into orbital angular momentum analogous to the Einstein-de Haas effect. We have further demonstrated that the EdH effect is more pronounced than in a dipolar BEC and occurs in combination with a twisting motion, where individual spin components acquire excessive angular momenta of opposite directions which can become considerably larger than the total angular momentum induced by the EdH effect. We find that this twisting motion is caused by the dynamical anisotropic deformation of the phase-space distribution and therefore is an effect unique to a Fermi gas. The twisting motion can be reversed by tuning either an external magnetic field or an s-wave scattering length in such a manner that the total kinetic energy of the system decreases and we have explained this feature as a result of stability properties investigated in Ref.~\cite{Sogo2009}, where a dipolar Fermi gas will contract along the polarization axis with decreasing kinetic energy.
Our results show that a Fermi gas could be a better candidate for experimental observation of the EdH effect, as demagnetization is more pronounced, and the spatial distribution of a single spin component shows a very distinct spatial pattern associated with an excessive orbital angular momentum induced by the counter-rotating spin components. In addition, the sensitivity of the twisting motion to the scattering length may lead to novel methods of measuring such scattering lengths in Dysprosium or Erbium.

\begin{acknowledgments}
The authors would like to thank K.~Fujimoto for useful discussions. This work was supported by KAKENHI Grant No. JP26287088 from the Japan Society for the Promotion of Science, a Grant-in-Aid for Scientific Research on Innovative Areas ``Topological Materials Science’’ (KAKENHI Grant No. JP15H05855), and the Photon Frontier Network Program from MEXT of Japan. U.~E. acknowledges support from a postdoctoral fellowship of the Japan Society for Promotion of Science (JSPS).

\end{acknowledgments}

\bibliography{refs}
\newpage

\section{Supplemental Material}
\subsection{Dipole-dipole interactions in two dimensions}
In the usual three-dimensional case, the Hamiltonian for dipole-dipole interaction is given by
\begin{align}
\label{eq:sm1}
H_\text{DDI}=&\frac12\int d^3r d^3r'\sum_{klmn}\hat\psi_k^\dagger(\vec r)\hat\psi_l^\dagger(\vec r')\nonumber\\
&\times V_{klmn}^{3D}(\vec r-\vec r')\hat\psi_m(\vec r')\hat\psi_n(\vec r).
\end{align}
Here, the interaction potential for the case of free magnetization is
\begin{equation}
\label{eq:sm2}
V_{klmn}^{3D}(\vec r)=c_d\frac{r^2\vec\sigma_{kl}\cdot\vec\sigma_{mn}-3(\vec r\cdot\vec\sigma_{kl})(\vec r\cdot\vec\sigma_{mn})}{r^5},
\end{equation}
where $\vec\sigma$ denotes the Pauli matrices. With the identities $\sigma^\pm=\sigma^x\pm i\sigma^y$, Eq.~(\ref{eq:sm2}) becomes
\begin{align}
\label{eq:sm3}
V_{klmn}^{3D}(\vec r)&=\frac{c_d}{r^5}\left[\sigma_{kl}^+\sigma_{mn}^+\left(-\tfrac34(x-iy)^2\right)\right.\nonumber\\
&\left.+\sigma_{kl}^-\sigma_{mn}^-\left(-\tfrac34(x+iy)^2\right)\right.\nonumber\\
\left.+\left(\sigma_{kl}^z\sigma_{mn}^z\right.\right.&\left.\left.-\tfrac14\sigma_{kl}^+\sigma_{mn}^- -\tfrac14\sigma_{kl}^-\sigma_{mn}^+\right)\left(x^2+y^2-2z^2\right)\right.\nonumber\\
&\left.+\left(\sigma_{kl}^z\sigma_{mn}^+ +\sigma_{kl}^+\sigma_{mn}^z\right)\tfrac32\left(-xz+iyz\right)\right.\nonumber\\
&\left.+\left(\sigma_{kl}^z\sigma_{mn}^- +\sigma_{kl}^-\sigma_{mn}^z\right)\tfrac32\left(-xz-iyz\right)\right].
\end{align}
To achieve a 2D setup, we assume a very strong harmonic confinement in the $z$-direction, such that all particles occupy the harmonic oscillator ground state with respect to the $z$-axis. We then split the field operators in Eq.~(\ref{eq:sm1}) into a product of 2D field operators and the harmonic oscillator ground state wave function $\hat\psi_m(\vec r)=\hat\phi_m(x,y)\chi(z)$, where $\chi(z)=\frac{1}{(\pi l_z^2)^{1/4}}e^{-z^2/2l_z^2}$ and $l_z=\sqrt{\hbar/m\omega_z}$. We substitute this into the Hamiltonian Eq.~(\ref{eq:sm1}) and carry out the integration over $z$ and $z'$. The 2D version of the dipole-dipole interaction potential is therefore given by
\begin{align}
\label{eq:sm4}
V_{klmn}^{2D}(x,y)=\int dzdz'|\chi(z)|^2|\chi(z')|^2V_{klmn}^{3D}(x,y,z-z')\nonumber\\
=\int dzdz'\frac{1}{\pi l_z^2}e^{-z^2/l_z^2}e^{-z'^2/l_z^2}V_{klmn}^{3D}(x,y,z-z').
\end{align}
Due to the symmetry of $\chi(z)$, the two bottom rows of Eq.~(\ref{eq:sm3}) vanish. It remains to compute the integrals
\begin{align}
\label{eq:sm5}
I_1(\rho)&=\frac{1}{\pi l_z^2}\int dzdz'\frac{e^{-\tfrac{z^2}{l_z^2}}e^{-\tfrac{z'^2}{l_z^2}}(x^2+y^2-2(z-z')^2)}{[x^2+y^2+(z-z')^2]^{5/2}}\nonumber\\
&=\frac{\sqrt{2}}{\sqrt{\pi}l_z^5}e^{\rho^2/4lz^2}\left[\left(2l_z^2+\rho^2\right)K_0(\tfrac{\rho^2}{4l_z^2})-\rho^2K_1(\tfrac{\rho^2}{4l_z^2})\right],\\
\label{eq:sm6}
I_2(\rho)&=\frac{3}{4\pi l_z^2}\int dzdz'\frac{e^{-z^2/l_z^2}e^{-z'^2/l_z^2}}{[x^2+y^2+(z-z')^2]^{5/2}}\nonumber\\
&=\frac{\sqrt{2}e^{\tfrac{\rho^2}{4lz^2}}}{\sqrt{\pi}4l_z^5}\left[K_0(\tfrac{\rho^2}{4l_z^2})+\tfrac{2l_z^2-\rho^2}{\rho^2}K_1(\tfrac{\rho^2}{4l_z^2})\right],
\end{align}
where $\rho^2=x^2+y^2$, and $K_0$ and $K_1$ denote the modified Bessel functions of the zeroth order and the first order, respectively.
We obtain as a final result the dipole-dipole interaction potential for a 2D system
\begin{align}
\label{eq:sm7}
V_{klmn}(\vec r)=&\frac{c_d\sqrt{2}}{\sqrt{\pi}l_z^5}\left(V_1(r)\left[\sigma_{kl}^z\sigma_{mn}^z-\tfrac14\sigma_{kl}^+\sigma_{mn}^-\right.\right.\nonumber\\
&\left.\left. -\tfrac14\sigma_{kl}^-\sigma_{mn}^+\right]-\tfrac14(x-iy)^2V_2(r)\sigma_{kl}^+\sigma_{mn}^+\right.\nonumber\\
&\left. -\tfrac14(x+iy)^2V_2(r)\sigma_{kl}^-\sigma_{mn}^- \right),
\end{align}
where $V_1(\rho)=\sqrt{\pi/2} l_z^5 I_1(\rho)$ and $V_2(\rho)=4\sqrt{\pi/2}l_z^5I_2(\rho)$.

\subsection{Contact interaction in 2D}
A simpler calculation for the contact interaction term 
\begin{align}
\label{eq:sm8}
H_\text{c}=&\frac{4\pi\hbar^2a}{m}\int d^3r \hat\psi_\uparrow^\dagger(\vec r)\hat\psi_\downarrow^\dagger(\vec r)\hat\psi_\downarrow(\vec r)\hat\psi_\uparrow(\vec r)\nonumber\\
=&\frac{4\pi\hbar^2a}{m}\int dz |\chi(z)|^4\nonumber\\
&\times\int d^2r\hat\phi_\uparrow^\dagger(x,y)\hat\phi_\downarrow^\dagger(x,y)\hat\phi_\downarrow(x,y)\hat\phi_\uparrow(x,y)\nonumber\\
=&\frac{2\sqrt{2\pi}\hbar^2a}{l_z m}\int d^2r\hat\phi_\uparrow^\dagger(x,y)\hat\phi_\downarrow^\dagger(x,y)\hat\phi_\downarrow(x,y)\hat\phi_\uparrow(x,y)\nonumber\\
=&g_\text{2D}\int d^2r\hat\phi_\uparrow^\dagger(x,y)\hat\phi_\downarrow^\dagger(x,y)\hat\phi_\downarrow(x,y)\hat\phi_\uparrow(x,y)
\end{align}
provides us with the effective coupling constant $g_\text{2D}=2\sqrt{2\pi}\hbar^2a/l_z m$.

\subsection{Time evolution of orbital angular momentum}
For a 2D system, the angular momentum is only defined for the $z$-component such that in the phase-space picture the expectation value of the angular momentum operator $\hat L=\hat x\hat p_y-\hat y\hat p_x$ for a spin component is given by
\begin{equation}
\label{eq:sm9}
L_{mn}=\int\frac{d^2r d^2p}{(2\pi\hbar)^2}\left(xp_y-yp_x\right)W_{mn}(\vec r,\vec p).
\end{equation}
We calculate its time evolution by multiplying the collisionless Boltzmann-Vlasov equation
\begin{widetext}
\begin{align}
\label{eq:sm10}
\frac{d}{dt}W(\vec r,\vec p)=\left(-\frac{\vec p}{m}\cdot\nabla_x+m\omega^2\vec r\cdot\nabla_p\right)W(\vec r,\vec p)+\frac{1}{i\hbar}\left(\left[W(\vec r,\vec p),i\hbar p\sigma_z+U^a(\vec r)+U(\vec r)-\tilde U(\vec r,\vec p)\right]\right)\nonumber\\
-\frac12\left(\left\lbrace\nabla_p W(\vec r,\vec p),\cdot\nabla_r \left(U^a(\vec r)+U(\vec r)-\tilde U(\vec r,\vec p)\right)\right\rbrace+\left\lbrace\nabla_r W(\vec r,\vec p),\cdot\nabla_p\tilde U(\vec r,\vec p)\right\rbrace\right).
\end{align}
\end{widetext}
with $xp_y-yp_x$ and subsequent integration over phase space. The single-particle part is straightforward:
\begin{align}
\label{eq:sm11}
\int& \frac{d^2r d^2p}{(2\pi\hbar)^2} \left(xp_y-yp_x\right)\left(-\frac{\vec p}{m}\cdot\nabla_x\right)W(\vec r,\vec p)=0,\\
\label{eq:sm12}
\int& \frac{d^2r d^2p}{(2\pi\hbar)^2} \left(xp_y-yp_x\right)m\left(\omega_x^2x\partial_{p_x}+\omega_y^2y\partial_{p_y}\right)W(\vec r,\vec p)\nonumber\\
&=m\int \frac{d^2r d^2p}{(2\pi\hbar)^2}\left(\omega_x^2-\omega_y^2\right)xy W(\vec r,\vec p)\nonumber\\
&=0.
\end{align}
It would only be non-zero in an anisotropic trap.

 Before we carry on, we separate all $2\times 2$ matrices such as $W(\vec r,\vec p)$ into a scalar and vector part by decomposing them in terms of the unit matrix and the Pauli matrices. For a matrix $A$, this means
\begin{equation}
\label{eq:sm13}
A=A_0 \openone+\vec A\cdot\vec\sigma,
\end{equation}
where $A_0=\tfrac12(A_{\uparrow\uparrow}+A_{\downarrow\downarrow})$ and $\vec A=2\text{Tr}(\vec\sigma A)$.

For angular momentum, $L_0$ corresponds to the total orbital angular momentum of the system which is changed by the Einstein-de Haas effect. We find, that it is not affected by the commutators in Eq.~(\ref{eq:sm10}), but rather the next-order gradient terms. However, to gain insight into the twisting motion, we must look at $L_z=L_\uparrow-L_\downarrow$, which in leading order is induced by the commutator term in Eq.~(\ref{eq:sm10}). Neglecting the anti-commutator, we find 
\begin{align}
\label{eq:sm14}
\frac{d}{dt}L_z\approx &\sqrt{\frac{2}{\pi}}\frac{c_d}{4l_z^5}\int\frac{d^2rd^2r'd^2p}{(2\pi\hbar)^2}\left(xp_y-yp_x\right)V_2(\vec r-\vec r')\nonumber\\
&\times\left[(x-x')^2N_x(\vec r')W_y(\vec r,\vec p)\right.\nonumber\\
\left.-(x-x')\right.&\left.(y-y')\left(N_y(\vec r')W_y(\vec r,\vec p)-N_x(\vec r')W_x(\vec r,\vec p)\right)\right.\nonumber\\
&\left.-(y-y')^2N_y(\vec r')W_x(\vec r,\vec p)\right],
\end{align} 
where $N_j(\vec r)=\int d^2p W_j(\vec r,\vec p)$. This expression only depends on transverse magnetization components $W_{x,y}$. Further, for isotropic phase space distributions, Eq.~(\ref{eq:sm14}) is zero. 

If we assume transverse magnetization to be small and for small times to have approximately the same phase-space distribution, such that $W_{x,y}(\vec r,\vec p)=M_{x,y}W_0(\vec r,\vec p)$, Eq.~(\ref{eq:sm14}) reduces to
\begin{align}
\label{eq:sm15}
\frac{d}{dt}L_z\approx & \frac{\sqrt2c_dM_xM_y}{\sqrt{\pi}4l_z^5}\nonumber\\
&\times\int\frac{d^2rd^2r'd^2p}{(2\pi\hbar)^2}\left(xp_y-yp_x\right)V_2(\vec r-\vec r')\nonumber\\
&\times\left[(x-x')^2-(y-y')^2\right]N_0(\vec r')W_0(\vec r,\vec p),
\end{align} 
where $N_0(\vec r')=\int d^2p'f(\vec r',\vec p')$, we can identify the two ingredients for the appearance of counter-rotating spin components: Non-zero magnetization in the x-y-plane ($M_x,M_y>0$) and an anisotropic deformation of the Fermi surface of the phase-space distributions for the transverse spin components. While the initial distribution fulfills neither condition, transverse magnetization does occur during the demagnetization dynamics. This in turn automatically leads to a Fermi surface deformation along the axis of transverse magnetization due to the anisotropy of DDI.
\end{document}